\documentclass[twocolumn]{aastex62}
\usepackage{amssymb,amsmath}
\usepackage{graphicx}
\usepackage[caption=false]{subfig}
\usepackage{booktabs}
\usepackage{multirow}
\usepackage[flushleft]{threeparttable}
\usepackage{natbib}
\usepackage{enumitem}

\newcommand{\degree}{^{\circ}}

\newcommand{\rosat}{{\it ROSAT}}

\newcommand{\xmm}{{\it XMM}}
\newcommand{\xmmnewton}{{\it XMM-Newton}}

\newcommand{\swift}{{\it Swift}}

\newcommand{\nustar}{\textit{NuSTAR}}

\newenvironment{putjo}{\noindent\ignorespaces\fontfamily{pcr}\selectfont\ignorespaces}{\ignorespacesafterend\,\,}

\accepted{May 16, 2020}
\submitjournal{\apj}
\shortauthors{Chalise et al.}
\begin{document}
	\title{Broad-band X-ray observation of broad-line radio galaxy 3C 109}
	\author{Sulov Chalise}
	\affil{Department of of Physics, Montana State University, P.O. Box 173840, Bozeman, MT 59717-3840, USA}
	\email{sulov.chalise@montana.edu}
	\author{Anne M. Lohfink}
	\affil{Department of of Physics, Montana State University, P.O. Box 173840, Bozeman, MT 59717-3840, USA}
	\author{Erin Kara}
	\affil{Department of of Physics, Massachusetts Institute of Technology, 77 Massachusetts Avenue, Cambridge, MA 02139-4307, USA}
	\author{Andrew C. Fabian}
	\affil{Institute of Astronomy, University of Cambridge, Madingley Road, Cambridge, CB3 0HA, UK}
	\begin{abstract}
		
		We present a study of the central engine in the broad-line radio galaxy 3C\,109. To investigate the immediate surrounding of this accreting, supermassive black hole, we perform a multi-epoch broad-band spectral analysis of a joint \nustar/\xmm\, observation (2017), an archival \xmm\, observation (2005) and the 105-month averaged \swift-BAT data. We are able to clearly separate the spectrum into a primary continuum, neutral and ionized absorption, and a reflection component. The photon index of the primary continuum has changed since 2005 ($\Gamma = 1.61 \substack{+0.02 \\ -0.01} \rightarrow 1.54 \pm{0.02}$), while other components remain unchanged, indicative of minimal geometric changes to the central engine. We constrain the high-energy cutoff of 3C\,109 (E$_{\text{cut}}= 49 \substack{+7 \\ -5}$\,keV ) for the first time. The reflector is found to be ionized (log $\xi$ = $2.3 \substack{+0.1 \\ -0.2}$) but no relativistic blurring is required by the data. SED analysis confirms the super-Eddington nature of 3C\,109 initially ($\lambda_{Edd} >$ 2.09). However, we do not find any evidence for strong reflection (R = $0.18 \substack{+0.04 \\ -0.03}$) or a steep power law index, as expected from a super-Eddington source. This puts the existing virial mass estimate of 2 $\times 10^{8}$M$_{\sun}$ into question. We explore additional ways of estimating the Eddington ratio, some of which we find to be inconsistent with our initial SED estimate. We obtain a new black hole mass estimate of 9.3 $\times 10^{8}$M$_{\sun}$, which brings all Eddington ratio estimates into agreement and does not require 3C\,109 to be super-Eddington.
		
	\end{abstract}
	
	
	\section{Introduction} \label{intro}
	Active Galactic Nuclei (AGN) are powered by accretion onto the central supermassive black hole (SMBH). The accretion system is made up of several parts: a central set of relativistic electrons called the corona, an accretion disk and an axisymmetric dusty torus \citep[and references therein]{heck2014}. The AGN accretion disk emits mostly in the optical/UV band. The X-rays are generated in the AGN corona, which Compton up-scatters the thermal accretion disk photons \citep{haardt1991,haardt}. The relativistic jets, powered by the SMBH, seen in some AGN produce additional high-energy and synchrotron radiation Thus, AGN are observable from the radio up to $\gamma$-rays.
	\par
	Although AGN are classified into several categories based on their observational properties, the AGN unification model explains most of this variation by a few physical differences such as inclination and luminosity \citep{urry2004}. A major factor that dictates the AGN's observational traits, but is not explained by the unification model, is the presence or absence of relativistic particle jets. About 10\% of AGN are observed with these jets that are ultra-luminous in the radio band \citep{katgert,kellerman2016}. These synchrotron radio jets inflate bubbles of relativistic plasma \citep{bubble} and dig their way through the interstellar medium out of the host galaxy creating huge lobes. This may result in the ejection or heating of the interstellar gas, which suppresses the SMBH accretion and ceases star formation \citep{jets}. Although it is safe to assume that some intrinsic properties of the AGN central engine will determine the presence of a strong jet component, the exact mechanism is still unknown. The black hole spin is considered to be a major factor in the jet strength \citep{blandford1977,blandford1990,wilson1995}. However, the discovery of many Seyferts with large measured spins but without a strong jet suggests some additional factor \citep{reynolds2014}. Another hypothesis is that the jet production is governed by the Eddington-scaled low mass black hole accretion states \citep{blandford2018}. The low-luminosity AGN that contain weak steady jets might be in the low/hard state,and the high-luminosity AGN might be in the very high/unstable state but cycle between soft sub-state and hard jet producing state \citep{meierbook2012}. Thus, 
	understanding the accretion-powered and X-ray-rich central engine is vital to unravel the AGN jet production mechanism.
	\par In this context, X-ray studies of AGN can be considered a direct probe of the AGN's central engine since X-rays are created close to the SMBH and can penetrate large columns of obscuring material often present in AGN \citep{gandhi}. In addition to the primary X-ray continuum from the corona, a typical AGN X-ray spectrum also contains a reflection component generated from the reprocessing of the primary X-ray emission on the surface of the accretion disk or the torus. This X-ray reflection spectrum can reveal substantial information about the central engine of the AGN \citep{fabian1995,gandhi} and has been used to constrain the spin of black holes \citep{parker2014,reynolds2019}, the radius of the inner-disk region \citep{ghosh} and the temperature of the corona \citep{relxillecut,lohfink2015}. While soft X-rays can be obscured by neutral or ionized gas, reflection features in the hard X-rays such as the fluorescent Fe\,K$\alpha$ line \citep{fabian1989} and the ``Compton hump" feature around 10 -- 25 keV due to Compton reflection/scattering of X-rays from the disk or pc scale torus \citep{ross1999}, can be used to study even highly-obscured AGN \citep{cthick}.
	\par Most AGN whose central engines have been studied are non-jetted. Jetted AGN, in addition to being scarce, are sometimes oriented such that the jet emission overpowers their spectra. Due to the limited studies of jetted AGN, we cannot be sure whether there are any differences in the central engine between jetted and non-jetted AGN. Broad-line radio galaxies (BLRGs) are ideal objects to study in this context as they are oriented such that the central engine is directly observable but the jet is not in the direct line-of-sight. Although BLRGs share the presence of broad optical lines with radio-quiet type I AGN, BLRGs typically have harder power-laws, smaller Fe\,K$\alpha$ equivalent widths and weaker reflection features \citep{grandi,ball2007,sam2009,lohfink2013,kingbasejet}. High quality broad-band X-ray studies of several BLRGs have been performed after the launch of the \nustar\, observatory \citep{nustar}. Using two \nustar\, observations of BLRG 3C\,382, \citet{ball2014} found the presence of Comptonizing corona with very weak reflection features, compared to what is usually found in Seyfert galaxies. Similarly, Comptonization parameters for the BLRG 3C\,390.3 were constrained by \citet{lohfink2015} using broad-band X-ray data. Using radio/X-ray data, a link between the events in the jet emission and accretion disk has been studied in the BLRG 3C\,111 \citep{chatterjee} and the BLRG 3C\,120 \citep{lohfink2013,3c120nustar}. Studying more BLRGs will provide valuable data points needed to study the difference between jetted and non-jetted central engines. 
	\par 
	In this paper, we study the broad-band spectra of 3C\,109, a luminous BLRG at $z = 0.306$. 3C\,109 stands out as a BLRG as its Eddington ratio has been estimated to be super-Eddington, while other BLRGs have Eddington ratios of a few tens of per cent at maximum \citep{ursini,lohfink2015}. We aim to explore this atypical nature of 3C\,109 in detail.
	\par There have been several studies of 3C\,109 in different emission bands. Its supermassive black hole mass has been estimated to be $\log (M/M\sun) \sim 8.3$ \citep{mclure}. Previous works on 3C\,109 using \textit{ASCA} and \textit{VLBI} observations revealed the presence of an Fe\,K$\alpha$ line, excess neutral absorption and a radio jet of $\sim$ 280 kpc scale. Using \textit{VLBI} data the inclination of 3C\,109's jet was constrained to $35 \degree < i < 56 \degree$ \citep{vlbi}. Using \textit{ASCA} data, \citet{allen} found a broad Fe\,K$\alpha$ line with an intrinsic FWHM $\sim 120,000$ km s$^{-1}$. Using \textit{XMM} observations \citet{3c109xmm} found 3C\,109 to be accreting above the Eddington limit along with the presence of blurred ionized reflection and excess neutral and ionized absorption.
	\par Here, we present the results of a joint \xmm\, ($\sim$ 64\,ks) and \nustar\, ($\sim$ 140\,ks) observation. The sensitivity of \nustar\, in the hard X-rays allows us to study the reflection spectrum of 3C\,109 in unprecedented detail. The partially simultaneous \xmm\, data provide us with valuable soft X-ray coverage. We focus on the broadband spectral analysis of 3C\,109, including the full reflection spectrum for the first time. This paper is structured as follows. In Section \ref{dred}, we report on the observations and data reduction. In Section \ref{spev}, we perform the short and long term spectral variability analysis. In Section \ref{span}, we present the analysis of the spectra. The results are discussed in Section \ref{dis}.
	\section{Data Reduction}
	\label{dred}
	\subsection{Overview}
	In this paper, we analyze the \textit{NuSTAR} \citep{nustar} observations of 3C\,109 along with a simultaneous \textit{XMM} pointing. In addition to these unpublished observations, we also consider an archival \textit{XMM} \citep{3c109xmm} observation and the 105-month averaged \swift\,-\textit{BAT} spectrum \citep{105bat}.  Table~\ref{obs} provides an overview of the X-ray observations considered and Table~\ref{obs_multi} of the observations at other wavelengths. 
	\begin{table}
		\caption{X-ray observations investigated in this paper.}
		\begin{tabular}{c|c|c}
			Observatory (ObsID) & Start time & Net Exposure \\
			\hline \hline \textit{NuSTAR} (60301011002) & 2017-08-20 01:56:09 & 61.4 ks\\
			\textit{NuSTAR} (60301011004) & 2017-08-22 20:06:09 & 81.7 ks\\
			\textit{XMM} (0200910101) & 2005-02-03 17:33:35 & 40.5 ks\\
			\textit{XMM} (0795600101) & 2017-08-20 02:33:43 &  63.6 ks\\
		\end{tabular}
	\end{table}\label{obs}
	\begin{table}
		\caption{\textit{XMM}-OM observations studied in this paper.}
		\begin{tabular}{c|c|c}
			OM Band (ObsID) & Start time & Net Exposure \\
			\hline \hline 
			W1 (0200910101) & 2005-02-03 17:33:35 & 39.2 ks\\
			W1, V (0795600101) & 2017-08-20 02:33:43 & 62.3 ks\\
		\end{tabular}
	\end{table}\label{obs_multi}
	\subsection{XMM-Newton}
	In this work we consider the data from two \textit{XMM} observations (Table~\ref{obs}). Data from both EPIC-pn and EPIC-MOS are considered. We have processed the \xmm\, data using the \textit{XMM-Newton} Scientific Analysis System (SAS version 15.0.0). The signal-to-noise of RGS was very low and no spectral information could be obtained. The EPIC data were first screened and periods of high particle backgrounds rejected. The spectra were produced from the created event files using \texttt{evselect}. Responses were created with \texttt{arfgen} and \texttt{rmfgen}. 
	
	The OM photometry was obtained using \texttt{omichain} and \texttt{omdetect}. The OM data were only of good quality in the W1 and V filters, all other filter bands were too affected by imaging artifacts to yield reliable photometry. As 3C\,109 is an elliptical galaxy, we assume that the galaxy contribution to the OM bands would be minimal.
	\subsection{NuSTAR}
	We produce cleaned event files using \texttt{nupipeline}. To reduce the times with high particle background but maximize exposure time, we study the background filtering reports provided by the \textit{NuSTAR} team. To minimize the effect of the SAA and thereby the background noise, we select SAAMODE ``optimized" and also exclude the ``tentacle" region. From the cleaned event files we extract source spectra (70" circular region) and the background spectra (95" circular region). 3-25\,keV light curves are also extracted with a 2\,ks time resolution. 
	\section{Spectral Variability}
	\label{spev}
	\subsection{Short-term Flux Variability}
	\label{s-t}
\begin{figure}[!h]
		\centering 
		\includegraphics[width=0.45\textwidth]{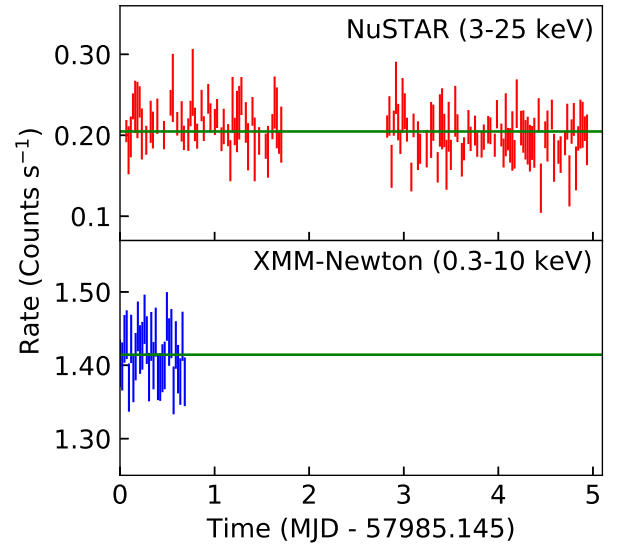}
		\caption{We show the background-subtracted \textit{NuSTAR} (top) and \textit{XMM}-Newton (bottom) light curves with 2 ks time resolution. A constant (green) is an acceptable fit for both light curves.}
		\label{short_term}
	\end{figure}

	We inspect the light curves (Fig \ref{short_term}) of the recent \nustar\, and \xmmnewton\, observations to assess the short-term spectral variability of 3C\,109. By eye, no variation is present, and a constant provides an acceptable description of the light curves (\nustar\,: $\chi^{2} = $ 125.2 for 121 d.o.f. ; \xmmnewton\,: $\chi^{2} = $ 22.8 for 29 d.o.f.). Without the presence of any short-term flux variability, we only consider the observation-averaged spectra for the remainder of this work. 
	\subsection{Long-term Flux Variability}
	\label{sec:longterm}
\begin{figure}[!h]
		\centering 
		\includegraphics[width=0.45\textwidth]{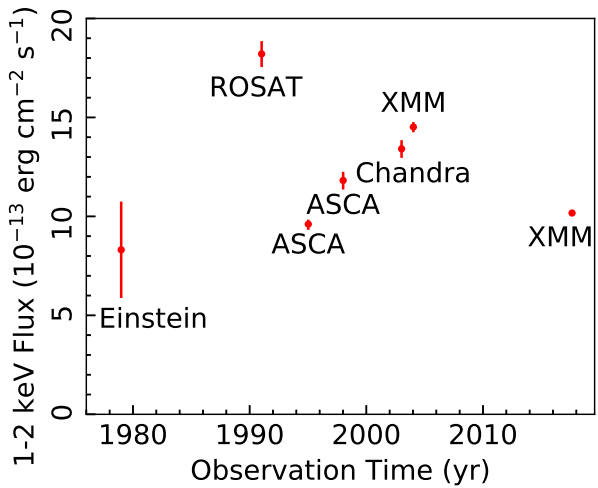}
		\caption{The historical X-ray flux of 3C 109 in the 1-2 keV band adopted from \citet{3c109xmm}. The telescope used to measure the flux is displayed along with each data point. The rightmost point is the recent flux value from \xmm. We see a 30\% flux decrease from the 2005 \xmm\, observation.}
		\label{long_term}
	\end{figure}

	We measure a flux of (10.2\,$\pm$\,0.1)$\times 10^{-13}$\,erg\,cm$^{-2}$\,s$^{-1}$ in the 1-2\,keV band using the \xmm\, \textit{EPIC-PN} observation and an absorbed power law model. \citet{3c109xmm} have compiled the long-term evolution of 3C\,109's flux in that energy band in Figure\,6 of their paper. Figure \ref{long_term} shows our new flux measurement along with those reported in \citet{3c109xmm}. Long-term X-ray variability is present with a 30\% flux decrease with respect to the last pointed X-ray observation of the source in 2005 and 17\% lower flux than the 39-year average.
	\section{Spectral Analysis}
	\label{span}
	X-ray spectral analysis is a great tool to study an AGN central engine. To constrain 3C\,109's central engine parameters, we perform a detailed analysis of the recent spectral \nustar\, and \xmm\, data in this section. We, then, continue with a `multi-epoch analysis' where we improve upon those constraints by including the archival \xmm\, data \citep{3c109xmm} and 105-month averaged \swift\,-\textit{BAT} data \citep{105bat}.
	\par Throughout this work, the spectral modeling is performed using \textit{XSPEC} 12.10.0c \citep{xspec}. We use the built-in $\chi^{2}$ minimization technique to fit the data, which are binned to a minimum of 20 counts per bin. All errors were calculated at 90\% confidence level. We always include a cross correlation constant in all of the subsequent model fits to account for any cross-calibration flux offsets among different detectors. Further, we include a galactic absorption component in all of our models to account for the ISM absorption from our galaxy. To describe the ISM absorption, we use the ISM absorption model \texttt{TBabs} \citep{tbgas} and fix the column density value to the total galactic column density, $N_{\text{H}} = 1.53\times 10^{21}$ cm$^{-2}$ obtained from the N$_{\text{H}}$ Tool \citep{nhtool}.
	\subsection{NuSTAR}
	\label{sec:nustar}
	As no short-term spectral variability was found (Section \ref{s-t}), an average spectra for each \nustar\, detector, FPMA and FPMB, are created from the two \nustar\, observations. The resulting spectra are analyzed in the energy range 3\,-\,70\,keV. The data above 70\,keV for \nustar\, are discarded due to high background. Both averaged spectra are fitted simultaneously. 
	\par
\begin{figure}[!h]
		\centering 
		\includegraphics[width=0.45\textwidth]{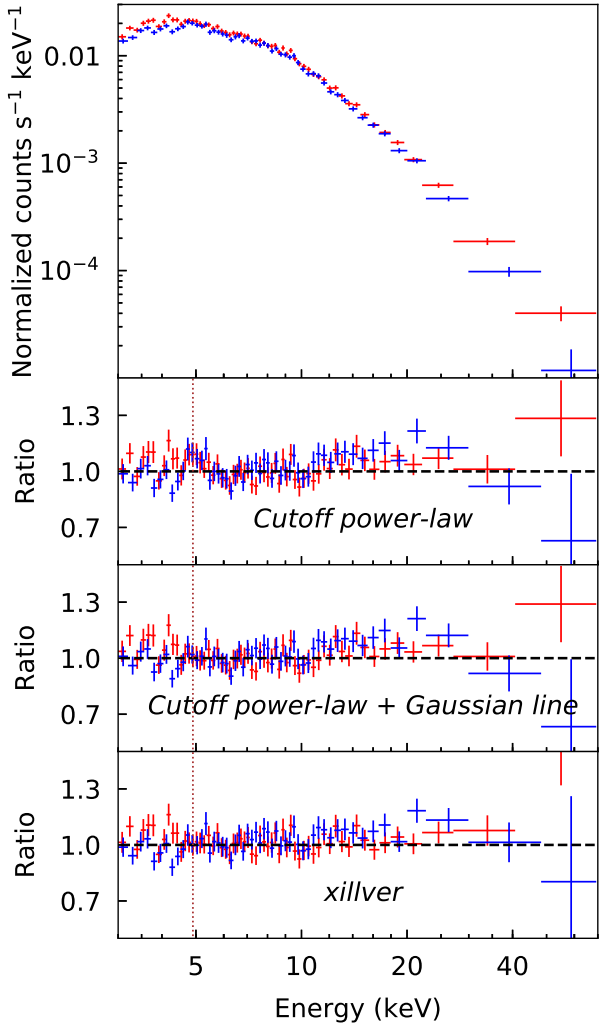}
		\caption{The counts spectrum (top) and the ratio residuals for the spectral modeling of the \textit{NuSTAR} data are shown for both FPMA (red) and FPMB (blue). The middle-top panel shows the ratio residuals for a cutoff power-law model. The middle-bottom panel shows the ratio residuals for the cutoff power-law model with a redshifted gaussian line. The bottom panel shows the ratio residuals for the self-consistent reflection model, \begin{putjo}xillver\end{putjo}. The dotted brown line marks the 6.4\,keV rest frame energy.}
		\label{nustar_res}
	\end{figure}

 To get a first glimpse of the spectral shape before performing any complex modeling, we initially fit a simple power-law model as a rough description of the primary X-ray continuum. We obtain an acceptable fit with $\chi^{2} = $ 1054.9 for 899 d.o.f. Now, we check for the presence of a high-energy cutoff to the power-law. We find that if we use a power-law model with a cutoff, the fit improves significantly ($\Delta \chi ^{2}$ improvement of 64.2 for 1 extra parameter). From a visual inspection of the ratio residuals of this fit around 5\,keV (Fig \ref{nustar_res},middle-top), we suspect the presence of an iron line. To model the probable iron line, we add a redshifted Gaussian line component to the cutoff power-law model. We fix the redshift to the known value for 3C\,109, z$= 0.306$. We obtain a $\Delta \chi ^{2}$ improvement of 17.1 for 3 extra parameters and a line at energy 6.47$\substack{+0.19 \\ -0.17}$\,keV with line width of 0.24$\substack{+0.30 \\ -0.24}$\,keV. The ratio residuals for this fit (Fig \ref{nustar_res},middle-bottom) show that the inclusion of the Gaussian line has removed the line-like feature from the previous-fit ratio residuals (positive ratio residuals around 5\,keV). With the detection of the high-energy cutoff and the iron line, we can confidently rule out a strong jet contribution in the X-ray band. 
	\par Although the previous model describes the averaged spectral data well, it is not a self-consistent model and the origin of the fitted line is unclear. Thus, we are motivated to use a self-consistent model that can naturally explain the line. We test the reflection model, \begin{putjo}xillver\end{putjo} \citep{xillver}, which explains the line as being caused by X-ray reflection of power-law photons off the accretion disk or other surrounding material. We assume a neutral reflector initially (log $\xi = 0$). We fix the reflector inclination to 35$\degree$, which is the upper limit from the VLBI observation for the jet inclination \citep{vlbi}. The power-law photon index ($\Gamma$), iron abundance ($\mathrm{Fe}_\mathrm{abund}$), high-energy cutoff ($E_\mathrm{cut}$) and reflection fraction ($R$) are free to vary. We obtain a good description of the spectra with $\chi^{2} = $ 960.9 for 895 d.o.f. and flat ratio (Fig \ref{nustar_res},bottom). We find a power-law photon index of $1.61 \substack{+0.12 \\ -0.08}$, an iron abundance of $2.1 \substack{+2.0 \\ -1.4}$, a high-energy cutoff of 55$\substack{+27 \\ -10}$ keV and a reflection fraction of $0.27 \pm 0.06$.
	\par Allowing for an ionized reflector (log $\xi > 0$) or relativistic blurring of the reflection using \begin{putjo}relxill\end{putjo} \citep{relxill} does not lead to a statistically significant fit improvement.  
	\subsection{XMM-Newton}
	\label{sec:xmm}
	After exploring the \nustar\, spectra, and clarifying the origin of the X-ray emission, we are ready to analyze the recent \xmmnewton\, spectra of 3C\,109. Spectra from all three EPIC CCD detectors are fitted simultaneously. As these detectors are sensitive, calibrated, and match in the energy range 0.3\,-\,10 keV for our observations, we discard the data points outside this range. Just as we did during the \textit{NuSTAR} analysis, we start again with a simple power-law model. This is a bad description of the data ($\chi^{2} = 12765$ for 2061 d.o.f.) This could be indicative of unmodeled absorption as excess obscuration of 3C\,109 along the line of sight modifies the spectrum below 10\,keV. Absorption could therefore explain the bad fit of the \textit{XMM} spectra but not the \textit{NuSTAR} spectra for the same model.
	\par
\begin{figure}[!h]
		\centering 
		\includegraphics[width=0.45\textwidth]{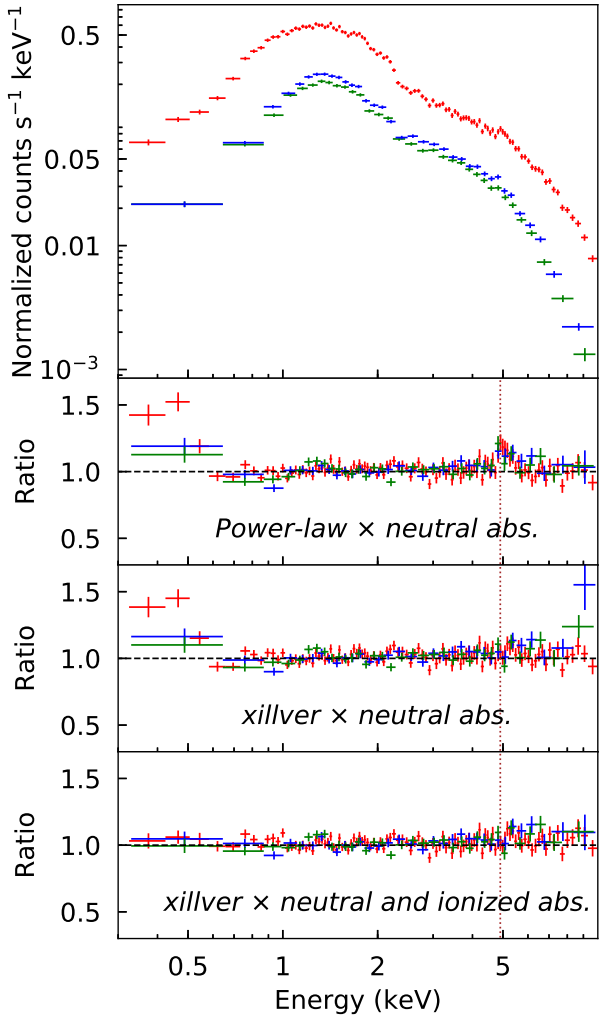}
		\caption{The counts spectrum (top) and the ratio residuals of the spectral modeling of the \xmm\, data are shown for \textit{EPIC-PN} (red), \textit{MOS-1} (blue) and \textit{MOS-2} (green). The middle-top panel shows the ratio residuals for a power-law model with intrinsic neutral absorption. The middle-bottom panel shows the ratio residuals for the self-consistent reflection model, \begin{putjo}xillver\end{putjo} with intrinsic neutral absorption. The bottom panel shows the ratio residuals for \begin{putjo}xillver\end{putjo} with intrinsic neutral and ionized absorption. The dotted brown line marks the 6.4\,keV rest frame energy.}
		\label{xmm_res}
	\end{figure}

  To improve the fit, we add a neutral absorption component to the power-law model. This component is modeled by \begin{putjo}zTBabs\end{putjo} with a free-to-vary column density (N$_{\text{H}}$) at the source's redshift in order to account for any intrinsic neutral absorption. This modification significantly improves the fit to $\chi^{2} = 2304.2$ for 2059 d.o.f. The ratio residuals of this fit (Fig \ref{xmm_res}, middle-top) show an iron line-like feature around 5\,keV, which corresponds to 6.4\,keV in the source rest frame.
	\par To explain the origin of the Fe\,K$\alpha$ line, we use the self-consistent reflection model, \begin{putjo}xillver\end{putjo} \citep{xillver}, still retaining the \begin{putjo}zTBabs\end{putjo} component. We assume a neutral reflector initially (log $\xi = 0$). The inclination is again fixed at 35$\degree$. The power-law photon index ($\Gamma$), iron abundance and reflection fraction are free to vary. The high-energy cutoff cannot be constrained below 10\,keV, so it is fixed at the best fit value of 55\,keV obtained from the \nustar\, modeling. The fit improves to $\chi^{2} = 2288.9$ for 2058 d.o.f. Allowing for an ionized reflector (log $\xi > 0$), we obtain a $\Delta \chi ^{2}$ improvement of 29.2 for 1 extra parameter. With the detection of ionized reflection, we can assume this reflector to be an accretion disk. The ratio residuals of this fit (Fig \ref{xmm_res}, middle-bottom) still show signs of extra unmodeled absorption around 0.5\,keV. The extra absorption was also seen by \citet{3c109xmm} and was attributed to the presence of an intrinsic ionized absorber. To model the ionized absorption, we create a custom photoionized absorption table model using the \begin{putjo}XSTAR2XSPEC\end{putjo} script, which uses the \begin{putjo}xstar\end{putjo}  \citep[v2.54 ;][]{xstar} code to produce an \begin{putjo}xspec\end{putjo}-compatible, multiplicative tabulated model grid. The model is calculated assuming the material has a covering fraction of unity, a typical temperature of 10$^{6}$\,K, a typical density of 10$^{12}$\,cm$^{-3}$, an ionizing luminosity of 3$\times$10$^{45}$\,ergs\,s$^{-1}$ (based on the observed flux), a turbulent velocity of 100\,km\,s$^{-1}$ and solar abundances. We also assumed a power-law spectrum with the photon index of 1.6 as illuminating flux. The resulting table model has two free parameters: the absorption column and the ionization parameter. The redshift of this absorber is fixed at the source redshift. We multiply this model component with the previous model, which leads to an improved fit with $\chi^{2} = 2114.1$ for 2055 d.o.f. The ratio residuals of this fit (Fig \ref{xmm_res},bottom) show no additional significant features. We obtain a power-law photon index of 1.55 $\pm{0.02}$, an iron abundance of $3.6 \substack{+6.4 \\ -1.7}$, an ionization parameter of 2.3$\substack{+0.1 \\ -0.2}$ erg cm s$^{-1}$ and a reflection fraction of $0.19 \pm 0.06$. These values are comparable to the values found in Section \ref{sec:nustar} for the modeling of the \nustar\, dataset. We are able to constrain the neutral absorbing column to $3.14 \pm 0.9 \times 10^{21} $ cm$^{-2}$, and the ionized absorbing column to $7.0 \pm 1.5 \times 10^{21} $ cm$^{-2}$ with ionization parameter of $1.02 \substack{+0.07 \\ -0.06}$ erg cm s$^{-1}$.
	\par Allowing for relativistic blurring of the reflection using \begin{putjo}relxill\end{putjo} \citep{relxill} does not lead to a statistically significant fit improvement.  
	\subsection{Multi-epoch analysis}
	\label{sec:multiepoch}
	As AGN have inherently variable spectra, we cannot average two spectra taken at different epochs. In particular, we have already established the long-term variability of 3C\,109 in Section \ref{sec:longterm}. So, we cannot utilize the 2005 \xmm\, observation to generate an averaged \xmm\, spectra with better signal-to-noise. However, we can perform a multi-epoch analysis, which utilizes data from different epochs, to improve constraints on the physical parameters expected to be non-variable between the epochs. Thus, we perform a multi-epoch analysis in which not only spectra from \textit{NuSTAR} and \textit{XMM} are fitted jointly but also spectra from different epochs are included.
	\par The 3C\,109 X-ray spectra included in this multi-epoch analysis are the recent \xmmnewton\, and \nustar\, spectra along with the archival 2005 \xmmnewton\, spectra \citep{3c109xmm} and the \swift-\textit{BAT} spectrum \citep{105bat}. The \swift-\textit{BAT} spectrum covers the very hard X-rays, and has a high signal-to-noise ratio as it is a long-term average spectrum. All spectra are fitted together by linking the parameters expected to remain constant, on physical grounds.
	\par We describe the spectra using the best fit model from the \xmm\, analysis, ionized reflection along with intrinsic neutral and ionized absorption. The parameter states, fixed or variable, are identical to Section~\ref{sec:xmm} except for the high-energy cutoff, which is free to vary due to the inclusion of hard X-ray data. All parameters are linked together among the different epochs except for the power-law photon index, reflection fraction and normalization. These parameters are linked between the recent \xmmnewton\, and \nustar\, spectrum. As \swift-\textit{BAT} only consists of hard X-ray data, it cannot constrain the reflection fraction and the high-energy cutoff independently and thus it is linked with the recent \xmmnewton\, and \nustar\ data, in order to able to obtain meaningful constraints.
	\par
\begin{figure}[!h]
		\centering 
		\includegraphics[width=0.49\textwidth]{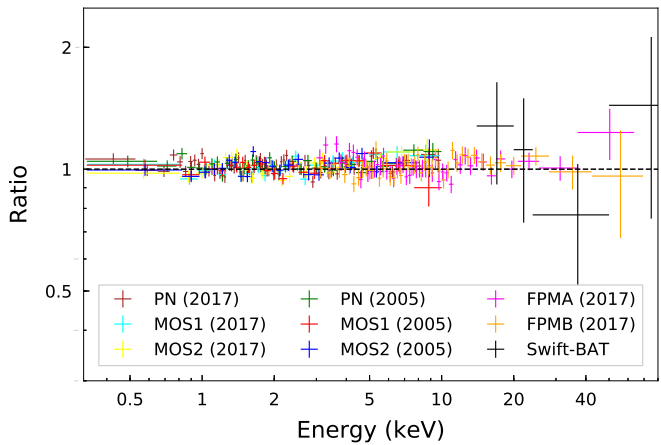}
		\caption{ The plot shows the best-fit ratio residuals for \xmm-PN (2017: brown, 2005: green),\xmm-MOS1 (2017: cyan, 2005: red),\xmm-MOS2 (2017: yellow, 2005: blue), \nustar\,-FPMA (magenta),  \nustar\,-FPMB (orange) and the 105-month averaged \swift-BAT data (black).}
		\label{mega_res}
	\end{figure}

 
	We find a neutral absorbing column of $3.38 \substack{+0.6 \\ -0.7}\times 10^{21} $ cm$^{-2}$, and an ionized absorbing column of $6.64 \substack{+1.0 \\ -0.9}\times 10^{21} $ cm$^{-2}$ with ionization parameter of $1.05 \substack{+0.07 \\ -0.05}$ erg cm s$^{-1}$. These match the values obtained in Section \ref{sec:xmm}. The power-law photon index appears to have changed between the two \xmm\, observations ($\Gamma = 1.61 \substack{+0.02 \\ -0.01} \rightarrow 1.54 \pm{0.02}$). The high-energy cutoff is constrained to $49 \substack{+7 \\ -5}$ keV, while the ionization of the reflector was found to be $2.3 \substack{+0.1 \\ -0.2}$  erg cm s$^{-1}$. The most-recent epoch parameter values are comparable to those of previous sections (\ref{sec:nustar}, \ref{sec:xmm}) but with improved constraints. In Table \ref{tab:bfit}, we report the best fit ($\chi^{2} = 4669.7$ for 4589 d.o.f.) values for all parameters of significance for this description of the data. In Figure \ref{mega_res}, we show the best-fit ratio residuals for \xmm-PN (2017: brown, 2005: green), \xmm-MOS1 (2017: cyan, 2005: red), \xmm-MOS2 (2017: yellow, 2005: blue), \nustar\,-FPMA (magenta),  \nustar\,-FPMB (orange) and the 105-month averaged \swift-BAT data (black). The model describes the data well, and allowing for relativistic blurring of the reflection using \begin{putjo}relxill\end{putjo} \citep{relxill} does not lead to a statistically significant fit improvement.
	\par The results are in agreement to what is found if only the recent \xmm\, and \nustar\, spectra are modeled together but with better constraints.
	\begin{table}[!t]
		\caption{Best fit parameter values of the multi-epoch Broad-band X-ray data.}
		\begin{threeparttable}
			\centering
			\label{tab:bfit}
			\begin{tabular*}{0.49\textwidth}{@{}lll@{}}
				\toprule
				\textbf{Parameter}                      & \textbf{Spectra$^{*}$} & \textbf{Value} \\ \midrule \midrule
				Neutral absorption column                 & All     & $3.38 \substack{+0.6 \\ -0.7}\times10^{21}$\,cm$^{-2}$              \\ \midrule
				Ionized absorption column                          & All                & $6.64 \substack{+1.0 \\ -0.9}\times10^{21}$\,cm$^{-2}$          \\ \midrule
				Ionized absorption ionization                          & All                & $1.05 \substack{+0.07 \\ -0.05}$\,erg cm s$^{-1}$            \\ \midrule
				\multirow{3}{*}{Power-law photon index} & 2017            & $1.54 \pm{0.02}$           \\
				& 2005               & $1.61 \substack{+0.02 \\ -0.01}$              \\ 
				& B					& $1.22 \substack{+1.2 \\ -0.2}$				\\\midrule
				Iron abundance                          & All                 & $4.0 \substack{+3.7 \\ -1.2}$             \\ \midrule
				High-energy cutoff                      & All                 & $49 \substack{+7 \\ -5}$ keV             \\ \midrule
				Ionization             &All             & $2.3 \substack{+0.1 \\ -0.2}$  erg cm s$^{-1}$           \\ \midrule
				\multirow{2}{*}{Reflection fraction}    & 2017+B            & $0.18 \substack{+0.04 \\ -0.03}$              \\
				& 2005              & $0.14 \pm 0.05$              \\ \bottomrule
			\end{tabular*}
			\begin{tablenotes}
				\small
				\item *  `2017' refers to recent \nustar\, and \xmmnewton\, spectra, `2005' refers to the archival \xmmnewton\, spectra and `B' refers to the time-averaged Swift-BAT spectra. `All' refers to all aforementioned spectra. For `All', the parameter values are linked among all spectra whereas, for `2017+BAT' they are linked between `2017' and `B'.
			\end{tablenotes}
		\end{threeparttable}
	\end{table}
	\section{Discussion}
	\label{dis}
	\subsection{Summary}
	\label{summary}
	In this paper, we present the analysis of the recent \nustar\, and \xmm\, observations of the broad-line radio galaxy 3C\,109 taken in 2017. We found the new X-ray data to have no significant short-term variability, but determined that the 1-2\,keV flux fluctuates on long timescales (years) as is expected for AGN \citep[][and references therein]{mchardy2010}. The 2-10\,keV unabsorbed flux was found to be 5.99$\pm 0.04 \times 10^{-12}$ erg cm$^{-2}$ s$^{-1}$ based on the \xmm\, observation and the best fit model, corresponding to a 2-10\,keV luminosity of 1.66$\pm 0.01 \times 10^{45}$ erg s$^{-1}$ in the source's rest frame. The spectral analysis of the recent observations has revealed that the X-ray spectra are well characterized by a primary X-ray continuum from the corona and its reflection from an ionized reflector. It is likely that this ionized reflector is the accretion disk. The observed soft X-ray spectrum is further modified by intrinsic neutral and ionized absorption.  This result agrees with the previous \xmm\,-based study by \citet{3c109xmm}, although we were unable to confirm evidence for a relativistic blurring of the reflection spectra using our model on the same observation. This discrepancy is likely due to a different model used, or an improvement on the \xmm\, data calibration.
	\par 
	We also perform a multi-epoch analysis using the recent 3C\,109 observations along with the available 2005 \xmm\, observation and the 105-month averaged \swift-\textit{BAT} spectra in order to obtain better constraints on the non-variable parameters. This also permits us to study the evolution of changeable AGN parameters within the full time-frame of the analysis. We find that there is a change of the photon index of the primary power-law continuum from 1.61$\substack{+0.02 \\ -0.01}$ to 1.54$\pm 0.02$ between the 2005 \textit{XMM} observation and the new 2017 observation, consistent with the softer-when-brighter behavior widely observed in AGN \citep{mchardy1999,papada}, while other parameters remain the same within errors. This might mean that 3C\,109's central engine has not changed significantly since 2005 and the photon index variation can be explained solely by the changes to the corona.
	\subsection{Contribution of the jet}
	\label{jetcont}
	3C\,109 is a FR II radio galaxy with two symmetric lobes with hot spots and core emission \citep{vlbi}. The core component of the jet emission can provide a flux contribution to the observed spectrum of 3C\,109. If this contribution is significant, spectral modeling must account for a jet component.
	\par As 3C\,109 is lobe-dominated, the majority of jet emission stems from the lobes. The size of 3C\,109's jet is 110 arcsec with a projected linear size of $\sim$ 280 kpc \citep{vlbi}. So, the core, the northern lobe and the southern lobe of the jet are well separated. X-ray and optical/UV observations of 3C\,109 only include the contribution from the weak jet core but not from the luminous lobes, as those are easily resolved.
	\par In Section \ref{sec:nustar}, we already ruled out a strong jet contribution to the X-ray band (i.e. coronal emission). However, we must check the jet contribution to 3C\,109's accretion disk spectrum in the optical/UV to be able to model it.
	\par
\begin{figure}[!h]
		\centering 
		\includegraphics[width=\linewidth]{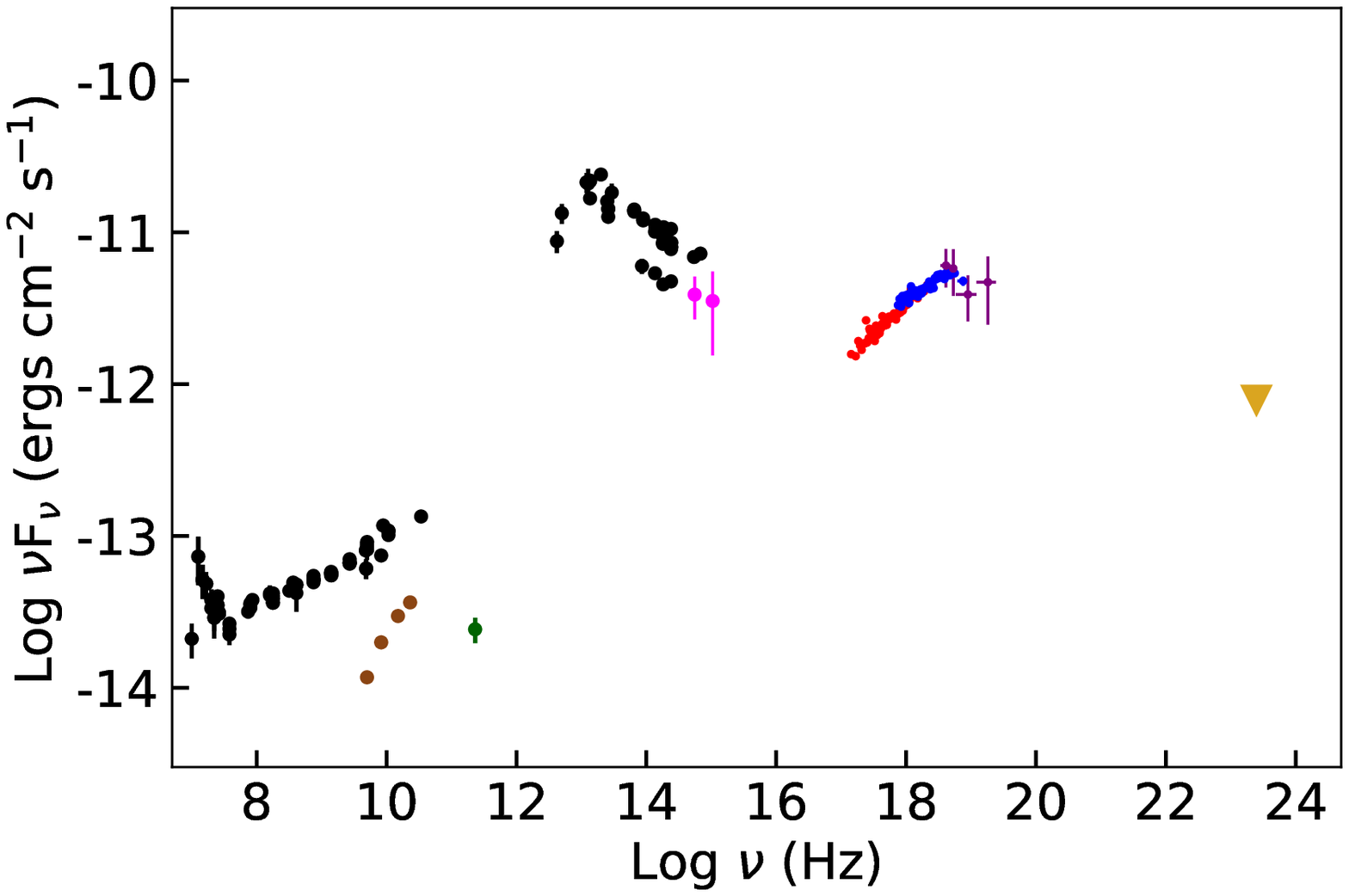}
		\caption{3C\,109 SED: The archival radio-optical flux measurements from the NASA/IPAC Extragalactic Database (NED) are shown in black. Brown points are the archival radio core flux from \citet{antonucci}. The green point is the flux measurement at 1300 microns from \citet{chini}. We plot the 2017 \xmm-OM flux (magenta). All the optical-UV flux are Milky-Way extinction corrected. We also plot the 2017 \xmm-PN (red), \nustar\,-FPMA (blue) along with the 105-month averaged \swift-BAT data (purple), all of which are absorption-corrected. }
		\label{ned_sed}
	\end{figure}

 In order to estimate this jet flux contribution, we study the SED of 3C109. Figure~\ref{ned_sed} shows the archival flux measurements in the radio-optical band from the NASA/IPAC Extragalactic Database (NED) as black points. At radio frequencies, the measurements suggest the presence of a strong radio-jet. However, these are the total radio flux (core+lobes) measurements. As an FR II radio galaxy, 3C\,109's jet has a well separated core, and only the core radio flux could be confused with the accretion disk emission. Thus, we plot the radio core flux (red points) as obtained by \citet{antonucci}, based on observations where the core and lobes were resolved. Also, \citet{chini} performed observation of 3C\,109 at 1300 micron (green point), and found that this 1300 micron flux measurement lies well below the extrapolation of the radio core spectrum using the core radio flux measurements from Antonucci et al. (1988).  They further noted that the 1300 micron flux measurement can be seen as an upper-limit of the high frequency spectrum of 3C\,109's radio core. This suggests that the radio core spectrum diminishes rapidly towards shorter wavelengths and makes a basically negligible flux contribution to the optical-X-ray band. These observations are some time in the past and one may wonder about the applicability of those results to our much more recent data. \citet{ekers} showed that the 5\,GHz radio core varies in the flux range 180--400 mJy from observations in the period 1974--1980. Subsequent 5\,GHz radio core flux measurements have also been within that range. Although, there is evidence of some variability, this variability is not strong enough to alter our results in any way.
	\par To give an overview of the energy budget of 3C109, we also include higher frequency data (2017 \xmm-OM: magenta ; 2017 \xmm-PN: red ; 2017 \nustar-FPMA: blue ; 105-month averaged \swift-BAT data: purple). All optical and UV fluxes shown are corrected for extinction from the Milky-Way, whereas all X-ray fluxes are corrected for both intrinsic and Milky-Way X-ray absorption. 3C\,109 is not detected in the $\gamma$-ray, and only its upper limit is shown in Figure \ref{ned_sed} (gold). The non-detection of 3C\,109 in the $\gamma$-ray regime is consistent with a weak jet core \citep{fermiblrg}.
	\par Hence, we can assume the X-ray and optical/UV emission in 3C\,109 is dominated by corona and accretion disk physics.
	\subsection{The lack of evidence of strong relativistic blurring}
	\citet{ball2007} concludes that the three necessary criteria for the production of powerful radio jets are: a rapidly spinning black hole, an inner accretion flow with a large H/r and and a favorable magnetic field geometry. A recent review paper by \citet{reynolds2019} confirms this. A consequence of the first criterion is the expected presence of strong relativistic blurring of the reflection spectra and ionized reflection. Even though 3C\,109 possesses a powerful radio jet, we find its spectra do not require the presence of relativistic blurring for the best fit model. This could be indicative of the lack of significant relativistic blurring effects of the reflection spectra. This lack may be explained by a truncated accretion disk, as it is predicted by the jet cycle model thought to operate in broad-line radio galaxies \citep{marscher2002,lohfink2013}. Modeling the recent data from Section \ref{sec:nustar} and \ref{sec:xmm} together with the relativistic reflection model, \begin{putjo}relxill\end{putjo} \citep{relxill}, we obtain a lower limit of 59 ISCO for the inner radius of the reflector assuming a maximally rotating black hole ($\chi^{2} = 3087.6$ for 2952 d.o.f.) However, the presence of a highly truncated accretion disk is at odds with our finding of a significantly ionized accretion disk with log $\xi$ = $2.3 \substack{+0.1 \\ -0.2}$.
	\par
	An alternative theory in which the corona is at a significant height from the accretion disk might explain this conundrum. This fits well with the idea that the AGN corona is the base of the particle jet, which can form at several tens of gravitational radii above the disk \citep{markoff2004,markoff2005,kingbasejet}. This height of the corona will minimize the relativistic effects near the ISCO while allowing the disk to be ionized. Modeling the multi-epoch data from Section \ref{sec:multiepoch} with the relativistic reflection model for a lamp-post geometry, \begin{putjo}relxilllp\end{putjo} \citep{relxill}, we obtain a lower limit of 336 gravitational radii for the coronal height assuming a maximally rotating black hole. We note that this model is a worse fit than the best-fit model ($\chi^{2} = 4711.5$ for 4585 d.o.f.) Further, the high value of the required height of the corona makes this theory a likely explanation for the lack of evidence of strong relativistic blurring.
	\par
	Another possible explanation for the absence of evidence of strong relativistic reflection is the presence of a corona in which the plasma is outflowing away from the disk. The outflow at mildly relativistic speeds causes aberration, reducing the X-ray emission towards the disk \citep{Beloborodov1999} and thus resulting in a reduced irradiation of the disk near the ISCO.
	\subsection{The UV/X-ray Spectral Energy Distribution}
	\label{sed-fit}
\begin{figure}[!h]
		\centering 
		\includegraphics[width=\linewidth]{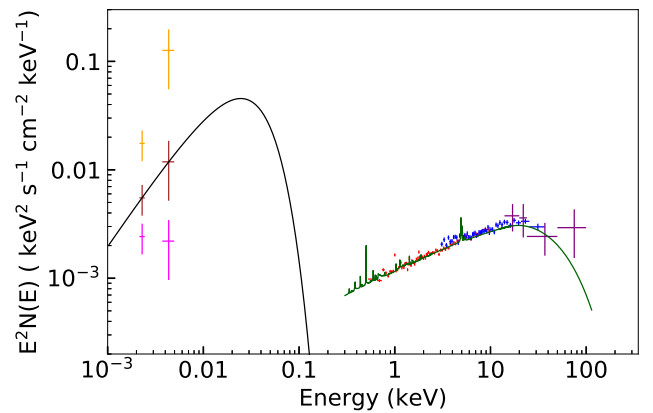}
		\caption{3C\,109 UV/X-ray SED: We plot the recent \xmm-PN (red), \nustar\,-FPMA (blue) along with the 105-month averaged \swift-BAT data (purple). All data are absorption-corrected. We also plot the Milky-Way extinction corrected \xmm-OM data (magenta) along with the source extinction correction using the previous E(B-V) value of 0.77 (orange) and the updated E(B-V) value of 0.32 (brown). The solid green line represents the unabsorbed X-ray reflection model and the solid black line represents the disk blackbody model assuming a maximum disk temperature of 10 eV.}
		\label{sed}
	\end{figure}

	In Figure \ref{sed}, we construct the UV/X-ray spectral energy distribution (SED) of 3C\,109 using the recent \xmm-PN (red), \xmm-OM (magenta), \nustar\,-FPMA (blue), along with the 105-month averaged \swift-BAT data (purple). We exclude the archival  \xmm-OM\, data as it has single data point in the W1 band, but we note that the observed archival W1 flux is 46\% higher than the recent W1 flux. The X-ray data and the best-fit X-ray model from Section \ref{sec:multiepoch} (green line) plotted in Figure \ref{sed} are corrected for both galactic and intrinsic absorption. The recent \xmm-OM (magenta) data is corrected for galactic extinction using reddening E(B-V) of 0.57. \citep{sch1998}.
	\par In Figure \ref{sed}, we also plot the accretion disk spectral model, \begin{putjo}diskpn\end{putjo} \citep{diskpn}, using a typical maximum disk temperature of 10 eV and an inner disk radius of 6 R$_{\mathrm{g}}$(= GM/c$^{2}$). This is shown as a black line in Figure \ref{sed-fit}. As we can clearly see, the \xmm-OM (magenta) data do not match the slope of this disk spectrum. Assuming that it is in fact the disk emission, this indicates further extinction of the \xmm-OM data.  
	\par Using infrared spectrophotometry, \citet{rudy1999} noticed that 3C\,109's infrared spectrum had significantly higher spectral index ($\alpha \sim 1.5$) than expected ($\alpha = -0.8$) from an accretion disk spectra. They use an accretion disk spectral model similar to ours, and calculate 3C\,109's in-situ reddening E(B-V) of 0.77. If we correct our \xmm-OM (magenta) data for an in-situ reddening, E(B-V) of 0.77 reported by \citet{rudy1999}, we find that these updated flux values (orange) also do not match an accretion disk spectrum. If we assume an in-situ reddening E(B-V) of 0.32, however, the resulting flux points (brown) do match an accretion disk spectrum. Given the time since the data analyzed in \citet{rudy1999} was taken, a change in reddening would not be unusual. We assume this new reddening to be true, and normalize the theoretical accretion disk spectrum such that it passes through the new flux points (brown).
	\par
	Additionally, we can relate the optical/UV accretion disk emission to the coronal X-ray emission using the spectral index, which is defined as $\alpha_{\text{ox}}$ = $\log$($ \text{L}_{2\,\text{keV}}$/$\text{L}_{2500 \text{\AA}}$)/ $\log$($ \nu_{2\,\text{keV}}$/$\nu_{2500 \text{\AA}}$), where $ \text{L}_{2\,\text{keV}}$ and $\text{L}_{2500 \text{\AA}}$ are the monochromatic luminosities 2 keV and 2500\,\AA\,respectively \citep{avni1982}. For 3C\,109, we find $\alpha_{\text{ox}}$ to be 1.33 $\pm{\,0.08}$. We have shown in Section \ref{jetcont}, 3C\,109 does not have a significant jet contribution to its optical and X-ray spectrum, and thus should have comparable $\alpha_{\text{ox}}$ to its radio-quiet counterparts. Using a sample of 545 X-ray selected radio-quiet type 1 AGN, \citet{lusso2010} found the mean $\alpha_{\text{ox}}$ of 1.37 $\pm{\,0.01}$. The 3C\,109 $\alpha_{\text{ox}}$ is only slightly deviated from this mean value.  Also, this $\alpha_{\text{ox}}$ agrees with the average $\alpha_{\text{ox}}$ = 1.44$\pm{\,0.13}$, \citet{green1995} obtained from the radio-loud sources from \rosat\, observations of The Large Bright Quasar Survey. For the radio-quiet sources of the same sample, they found $\alpha_{\text{ox}}$ of 1.57 $\pm{\,0.15}$. However, this study does not account for any contribution of jet to the optical and X-ray band, and should be interpreted with caution. Using $\gamma$-ray emission from 18 BLRGs, \citet{fermiblrg} showed that for these BLRGs the non-thermal (jet) component can contribute a few percent up to 35 percent of the observed SED. So, without properly accounting for jet contribution, the $\alpha_{\text{ox}}$ mean for a sample of radio-loud AGN is unreliable.
	\subsection{Super-Eddington outlier or faulty SMBH mass estimate?}
	\label{seout}
	Using the best-fit SED model from Section \ref{sed-fit}, we find the lower limit on the 3C\,109's bolometric luminosity to be 5.2$\times 10^{46}$ erg s$^{-1}$. Utilizing the black hole mass of $\log (M/M_{\sun}) = 8.3 \pm {0.4}$ from an H$\beta$ virial mass estimation \citep{mclure}, we find a lower limit on the Eddington ratio of 2.09. Similarly, using a bolometric correction of 20 \citep{vasudevan2007}, we find a bolometric luminosity estimate of $L_{bol} \sim 3.32 \times 10^{46}$ erg s$^{-1}$ from the 2-10\,keV luminosity corresponding to an Eddington ratio of 1.32$\substack{+2.0 \\ -0.8}$. Both of these Eddington ratio estimates appear to agree with the super-Eddington nature of 3C\,109 as reported by \citet{3c109xmm}.
	\par  Although observed to have a seemingly super-Eddington luminosity, 3C\,109 lacks other characteristics of super-Eddington AGN. Super-Eddington sources usually exhibit softer power-law spectral indexes ($\Gamma>2.5$), strong reflection and often possess strong outflows \citep{laurent,zubovas}. As 3C\,109 lacks all three of the expected features, we question its super-Eddington nature. 
	\par We note that both Eddington ratio estimates depend on the SMBH mass estimate. Additionally, the estimate from the SED modeling is also dependent on the accretion disk maximum temperature, which we have fixed to a value of 10\,eV. Thus, we seek an additional way of estimating the Eddington ratio independent of the black hole mass.
	\par Using a sample of 92 bright, soft X-ray selected AGN, \citet{grupe2010} found a correlation between $\alpha_{\text{ox}}$ and Eddington ratio ($\lambda_{Edd} $) given by the relation,
	\begin{equation}
	\alpha_{\text{ox}} = (0.11 \pm 0.02) \log \lambda_{Edd} + (1.39 \pm 0.02)
	\label{correlation}
	\end{equation}
	Using our $\alpha_{\text{ox}}$ value of 1.33 $\pm{\,0.08}$, we find an Eddington ratio of 0.28$\substack{+1.43 \\ -0.06}$. Additionally, we note that as the radio-loud AGN have smaller $\alpha_{\text{ox}}$ values than radio-quiet AGN, this Eddington ratio could be an overestimate. 
	\par
\begin{figure}[!h]
		\centering 
		\includegraphics[width=\linewidth]{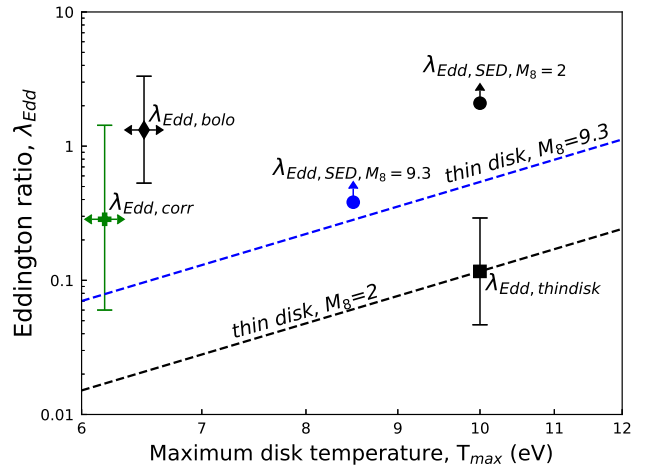}
		\caption{Eddington ratio vs Maximum disk temperature. The dashed lines are the theoretical prediction for a thin accretion disk using Equation \ref{theory} for SMBH mass of M$_8$=2 (black) and M$_8$=9.3 (blue). Eddington ratio estimate for a thin accretion disk assuming M$_8$=2 and T$_{\text{max}}$=10 eV, $\lambda_{\textit{Edd,thindisk}}$, is shown using a black `square'. The Eddington ratio estimate using the correlation from Equation \ref{correlation}, $\lambda_{\textit{Edd,corr}}$, is shown using a green `plus'. The Eddington ratio estimate using the bolometric correction factor of 20 on the 2-10\,keV luminosity of 3C\,109 with SMBH mass of 2$\times$10$^{8}$ M$_{\sun}$, $\lambda_{\textit{Edd,bolo}}$, is shown using a black `diamond'. The Eddington ratio lower limit from the SED fitting is shown using a `circle'. The black circle assumes a SMBH mass of 2$\times$10$^{8}$ M$_{\sun}$ and a maximum disk temperature of 10 eV, while the blue circle assumes a SMBH mass of 9.3$\times$10$^{8}$ M$_{\sun}$ and a maximum disk temperature of 8.5 eV.}
		\label{disktemp}
	\end{figure}

 In order to compare these three Eddington ratio estimates in relation to the parameters they depend on, we turn to a theoretical estimate. Assuming a non-rotating black hole with a thin accretion disk, Eddington ratio and the maximum accretion disk temperature in Kelvin, T$_{\text{max}}$, are related as 
	\begin{equation}
	\lambda_{Edd} \approx 0.583 \times \Big(\frac{1.8}{f_{col}}\Big)^{4} \times 10^{7} \text{M}_{8} \times \big(k\text{T}_{\text{max}}\big)^{4}
	\label{theory}
	\end{equation}
	\citep{thindisk} where, $f_{col}$ is the color temperature correction, M$_{8}$ is the black hole mass in units of 10$^{8}$ M$_{\sun}$ and $k$ is the Boltzmann constant in units of keV/Kelvin. Here, we use a color temperature correction of 1.8, a suitable value for accretion close to the Eddington limit \citep{shimura1995}. In Figure \ref{disktemp}, we plot the theoretical Eddington ratio estimate using Equation \ref{theory} for a range of maximum accretion disk temperature for the originally observed 3C\,109 mass of 2$\times$10$^{8}$ M$_{\sun}$ as a black dashed line. We also include the expected Eddington ratio for a maximum accretion disk temperature of 10\,eV. It is shown as a black `square' and annotated as $\lambda_{\textit{Edd,thindisk}}$. Its errorbars originate from the SMBH mass uncertainty.
	\par The observational Eddington ratio estimates are shown as well in Figure \ref{disktemp}. The Eddington ratio estimate using the lower limit on the bolometric luminosity from the SED fitting is shown using a black `circle' and annotated as $\lambda_{\textit{Edd,SED,M}_{\textit{8}}=2}$. Also, the Eddington ratio estimate using the bolometric correction factor of 20 on the 2-10\,keV luminosity is shown using a black `diamond' and annotated as $\lambda_{\textit{Edd,bolo}}$. Again the uncertainty originates from the SMBH mass estimation. Finally, the Eddington ratio estimate using the Equation \ref{correlation} is shown using a green `plus' and annotated as $\lambda_{\textit{Edd,corr}}$. Its errors are those of Equation \ref{correlation}.
	\par As we can clearly see in Figure \ref{disktemp}, the mass and temperature independent Eddington ratio estimate using Equation \ref{theory} does not agree with the Eddington ratio estimate using the lower limit on the bolometric luminosity from the SED fitting. Also, they both do not agree with the theoretical Eddington ratio estimate for a SMBH mass of 2$\times$10$^{8}$ M$_{\sun}$ and maximum accretion disk temperature of 10\,eV. Hence, we search for a new maximum accretion disk temperature and SMBH mass, which will unite all the Eddington ratio estimates.
	\par We begin by finding the SMBH mass that will make the Eddington ratio estimate using the bolometric correction factor of 20 equal to the mass and temperature independent Eddington ratio estimate derived from Equation \ref{correlation}. This mass is 9.3$\times$10$^{8}$ M$_{\sun}$ and is our new SMBH mass estimate. Now, we plot the theoretical Eddington ratio estimate for a range of maximum disk temperatures for this new mass, M$_8$=9.3, as a blue dashed line. We now find the maximum disk temperature that equates this new theoretical Eddington ratio estimate to the mass and temperature independent Eddington ratio estimate using Equation \ref{correlation}, i.e. find the T$_{\text{max}}$ where the $\lambda_{\textit{Edd,corr}}$ measurement meets the blue line. This maximum disk temperature is 8.5\,eV and is our new maximum disk temperature estimate. Now, using these new estimates, we find a new lower limit on the Eddington ratio from the SED fitting, which is shown using a blue `circle' and annotated as $\lambda_{\textit{Edd,SED,M}_{\textit{8}}=9.3}$.
	\par Thus, if we assume a SMBH mass of 9.3$\times $10$^{8}$ M$_{\sun}$ and a maximum disk temperature of 8.5 eV for 3C\,109, all of the Eddington ratio estimates agree with a value of 0.28. If correct, 3C\,109 would not be super-Eddington, consistent with the lack of the expected super-Eddington features in the spectrum. We caution that this new mass estimate has several assumptions/simplifications associated with it, which affect its reliability. Hence, an updated 3C\,109 mass estimate using more accurate and less error prone techniques, such as reverberation mapping, is needed before we can classify 3C\,109 as a sub-Eddington AGN with absolute certainty. 
	\subsection{Low temperature corona}
	\citet{ricci2017} found a median high-energy cutoff energy of $381 \pm 16$\,keV for 731 non-blazar AGN from a sample of 836 \swift /BAT objects with soft X-ray observations. Meanwhile, the best-fit high-energy cutoff of 3C\,109 is only $49 \substack{+7 \\ -5}$\,keV (Table \ref{tab:bfit}) which is much lower than average non-blazar AGN value. To put this in physical context, we can approximate the electron temperature of every optically-thin plasma to be half of the high-energy cutoff \citep{pet2001}. Thus, 3C\,109 coronal electron temperature is approximately $25 \pm 3$\,keV. This low electron temperature can be explained by the strong radiation fields expected from a super-Eddington disk, which cools the corona via Compton cooling \citep{kara2017}. However, this explanation is contingent on 3C\,109 being super-Eddington.
	\par
	Alternatively, the presence of hybrid plasma consisting of both thermal and non-thermal particles can also explain the observed low cutoff energy \citep{fabian2017}. The coronae harboring hybrid plasma are compact and highly magnetized. The compactness of an AGN corona can be quantified as a dimensionless radiative compactness parameter \citep{fabian2015}: 
	\begin{equation}
	\ell = \frac{\text{L} \cdot \sigma_{\text{T}}}{\text{R} \cdot m_{e}c^{3}}
	\label{ell}
	\end{equation}\\
	where, L is the luminosity, R is the radius, $\sigma_{\text{T}}$ the Thomson cross section and $m_{e}$ the mass of the electron. This radiative compactness will modulate the electron-positron pair production which in turn will regulate the coronal temperature \citep{fabian2015}. For 3C\,109, we do not have a coronal size estimate. Thus, we assume a spherical corona, 10 gravitational radii in size which is a conservative assumption as pointed out by \citet{fabian2015}. Using the mass of 2$\times$10$^{8}$ M$_{\sun}$, we get R$=2.9\times10^{14}$cm. From the 0.1-200\,keV unabsorbed X-ray continuum luminosity (6.0$\pm 0.2 \times$10$^{45}$ ergs/s) calculated from the joint 2017 \xmm-\nustar\, fit, we find $\ell$ = $550 \pm 10$ accounting just the uncertainty in luminosity. This value is higher than the usual range between 10 and 100 \citep{fabian2015}.  Coupled with the low coronal temperature, it is evident from Figure 1 of \citet{fabian2017} which presents the Electron temperature – compactness distribution along with the runaway pair production boundaries, the thermal pair-production as a dominating cooling mechanism for 3C\,109 is questionable.  Although 3C\,109 lies on the allowed region of this plot, it is not clear what stabilizes this thermal coronae lying well below the pair production limit. \citet{fabian2017} also show that even a small fraction of non-thermal particles in compact corona leads to lower equilibrium temperatures. A nonthermal fraction between 0.23 and 0.29 yields a stable corona (Figure 3 (left panel) of \citet{fabian2017}). Thus, a hybridized corona can also explain the low coronal temperature of 3C\,109. But, sensitive hard X-ray observation is needed to confirm this hypothesis \citep{fabian2017}. 
	\par Although, if we assume the new SMBH mass estimate from section \ref{seout}, we find $\ell$ of $118 \pm 10$ from Equation \ref{ell}. This new value is still well below the pair production limit. Thus retaining the possibility of hybridized plasma.
	\acknowledgments
	We thank the anonymous referee for his/her valuable suggestions and comments. This work was supported under NASA Contract No. 80NSSC17K0615, and made use of data from the NuSTAR mission, a project led by the California Institute of Technology, managed by the Jet Propulsion Laboratory, and funded by the National Aeronautics and Space Administration. This research has also made use of the NASA/IPAC Extragalactic Database (NED), which is funded by the National Aeronautics and Space Administration and operated by the California Institute of Technology."

	\vspace{5mm}
	\facilities{\nustar, \xmm, \swift }

\end{document}